# Mechanical Degradation of Unentangled Polymer Melts under Uniaxial Extensional Flow

Mingchao Wang[1], Stephen Sanderson[1] and Debra J. Searles[* 1, 2, 3]

[1]Centre for Theoretical and Computational Molecular Science, Australian Institute for Bioengineering and Nanotechnology, The University of Queensland, St Lucia, QLD 4072, Australia

[2]School of Chemistry and Molecular Biosciences, The University of Queensland, St Lucia, QLD 4072, Australia

[3]ARC Centre of Excellence for Green Electrochemical Transformation of Carbon Dioxide, The University of Queensland, St Lucia, QLD 4072, Australia

[*]Corresponding author: email: d.bernhardt@uq.edu.au

**Abstract**

Complex flow fields govern the deformation of polymers in various manufacturing processes. However, high flow rates may trigger reaction events (i.e., bond breaking or undesirable reaction of mechanophores) in raw polymeric materials, leading to the mechanical or functional debasement of manufactured structures. Additionally, it is difficult to fully characterize such molecular-level flow in the laboratory due to time- and length-scale limits. In this study, we perform non-equilibrium molecular dynamics (NEMD) simulations to explore the rheological and mechanical degradation of unentangled polymer melts under uniaxial extensional flow (UEF), allowing for chain breaking. Our simulations demonstrate shear thickening-thinning-thickening stages with the increase of UEF extension rates, resulting from flow-induced changes of chain conformation. With further increasing UEF extension rates, a bond-breaking potential leads to another flow thinning stage. Interestingly, fracture kinetics is originally first-order owing to the need for highly stretched polymer chains before bond fracture. It is no longer first-order when bond fracture is instigated before chains are stretched. Our computational work provides insight into the optimal design of the manufacturing process for polymeric materials.

# 1. Introduction

Polymer mechanochemistry explores the energy transformation between external mechanical forces and chemical reactions in polymers.[1, 2] Such mechano-chemical coupling has been successfully harnessed in developing novel polymeric materials for self-reporting[3, 4] and self-strengthening applications.[5, 6] To analyze this in the laboratory, small force-sensitive molecules (mechanophores) can be linked to polymer backbones to activate colorimetric and luminescent changes,[7, 8] as well as cross-linking reactions.[9, 10] Mechanophores have been incorporated into different states of polymers (i.e. dilute solutions, melts, and solids) to analyze the stress transfer from external loadings to mechanophores, and their reaction kinetics.[11] Recent experiments demonstrate that the activation of mechanophores strongly depends on many material factors, such as chemical environment and atomic structures of polymer matrices. As an example, the activation of spiropyran mechanophores in polymer solids is governed more by the polymeric network structures than their own molecular features.[12] The fundamental understanding of polymer mechanochemistry thus requires decoupling of these dominating factors.

Among various polymeric states, dilute solutions under ultrasonication are often selected for studying intrinsic mechano-chemical coupling.[13] The sonication-induced shear forces generate a velocity gradient along the polymer backbone, driving the effective activation of mechanophores.[13] However, the complicated loading conditions during ultrasonication can hinder the practical applications of such techniques in the general study of polymer mechanochemistry. Flow under controlled settings serves as an alternative way of exploring the individual effects of reaction parameters, such as chain lengths, solvent viscosities, and flow rates.[14] In addition, flow plays a significant role in the manufacturing processes of mechanophore-incorporated polymers.

The raw materials such as polymer solutions or melts undergo different types of flow (i.e. extension or contraction) during the inkjet printing or injection molding,[1] and manufacturing flows may trigger destructive reaction events in polymeric materials. Strong flow rates during processing can generate undesirable chain scissions or mechanophore activation, resulting in mechanical or functional alteration of manufactured materials.[1] As a result, it is necessary to explore the molecular deformation and failure of polymers under flows.

Given the mechano-chemical behavior originates at the molecular scale, flow is often occurring within enclosed apparatus and any measurement device will disrupt the flow, it is often difficult to experimentally characterize the molecular deformation and bond fracture in polymer chains under flow deformation. Alternatively, atomistic modelling has been widely conducted to successfully reveal the rheological properties of polymer systems under flows.[15] In general, extensional flow tends to lead to shear thickening (increase in viscosity) after a linear regime,[16-19] while planar shear flow tends lead to shear thinning (reduction in viscosity).[20-23] However, the use of solely rate-dependent rheological phenomena cannot provide a full understanding of the load transfer and molecular deformation in polymer chains, which also relies on their molecular structures, such as chain architectures and topologies. With an increase of shear strain rates, for example, it has been predicted that chain rotation starts dominating the molecular deformation of highly entangled polymer melts,[24, 25] rather than simple chain stretching in their unentangled states.[23] In addition, most atomistic simulations only consider finitely extensible bonds which still stay intact under intermediate- and high-rate flows.[26] This simulation assumption neglects the possibility of flow-induced bond fracture[27-29] or other chemical reactions (i.e. mechanophore activation)[30-32] in polymer chains, and is consequently incapable of predicting flow-dependent mechano-chemical behavior of polymer systems. Although a few theoretical models have been developed to simulate chain scission in dilute polymer solutions at low flow rates,[33, 34] there is still a lack of detailed investigations on the chain scission of concentrated polymer solutions or polymer melts under extensional flows, especially at high extension rates.

With the above in mind, here we carry out non-equilibrium molecular dynamics (NEMD) simulations of bead-spring polymer models to investigate the rheological behavior and chain scission of unentangled polymer melts with different chain lengths under uniaxial extensional flow (UEF). Two molecular potentials are utilized in the NEMD simulations: one that does not allow bond-breaking and another that does. With increase in the UEF extension rate, both potentials predict similar trends for the steady-state viscosities of unentangled polymer melts that follow the expected thickening-thinning-thickening stages. Such flow behavior corresponds to the flow-induced evolution of the chain conformations. With further increase in the UEF extension rate, the breakable-bond potential predicts another flow thinning stage due to the accumulation of broken bonds. Interestingly, kinetics is initially first-order owing to the significant stretching of polymer chains required before bond fracture. After some time, it is no longer first-order because bond fracture becomes faster than chain stretching.

## 2. Computational Methods

All molecular dynamics simulations are conducted on the basis of coarse-grained (CG) spring-bead models,[26] using the massively parallelized package LAMMPS.[35] All results are presented in Lennard-Jones reduced units. A self-avoiding random-walk approach[36] is utilized to construct all polymer melt models with an initial bead density of 0.85 consisting of 800 chains, each of which have lengths of $N = 20 – 80$ monomer beads. Since these chain lengths are shorter than the entanglement length of such spring-bead model ($N_e \approx 85$ beads), our polymer melt models are all in their unentangled states.[37] To obtain the initial equilibrium states of simulation models, the non-bonded inter-molecular and intra-molecular repulsive interactions among beads are simulated by a shifted and pure repulsive Weeks-Chandler-Andersen (WCA) potential,[38] which is a Lennard-Jones (LJ) potential with a cutoff radius of $R_c = 2^{1/6} \sigma$ given by:

$$U_{LJ}(r) = \begin{cases} 4\varepsilon\left[\left(\frac{\sigma}{r}\right)^{12} - \left(\frac{\sigma}{r}\right)^{6}\right] + \varepsilon & , r < R_c \\ 0 & , r \geq R_c \end{cases} \tag{1}$$

where $r$ is the inter-bead distance. This potential generates a pure repulsive force, and accounts for the effect of excluded volume. In the following, values of all variables will be given in Lennard-Jones reduced units.[39] The bond interactions between neighboring beads are modelled by the finitely extensible nonlinear elastic (FENE) potential:[26]

$$U_{FENE}(r) = \begin{cases} -\frac{1}{2}KR_0^2 \ln\left[1 - \left(\frac{r}{R_0}\right)^2\right] & , r < R_o \\ \infty & , r \geq R_o \end{cases} \tag{2}$$

where $K = 30$ is the spring constant; $R_0 = 1.5$ is the maximum bond length. Polymer melt models simulated by the combination of the WCA and FENE potentials (LJ+FENE) were equilibrated in $NVT$ ensembles by both Langevin and Nosé-Hoover atomic thermostats,[40, 41] at a temperature of $T = 1$ for a duration of $2.5\times10^5$ time units with a time step of $\Delta t = 0.005$. It was found that equilibrium results are independent of thermostat type used.

After the initial equilibration, a quartic potential was used to replace the FENE potential to describe a bond interaction between neighboring beads, which allows for their possible irreversible breakage under flow,[27-29]

$$U_Q(r) = \begin{cases} K_Q(r - R_Q)^2(r - R_Q - B_1)(r - R_Q - B_2) + U_0 & , r < R_Q \\ 0 & , r \geq R_Q \end{cases} \quad (3)$$

where $K_Q = 1200$ is the bond stiffness; $R_Q = 1.3$ is the cutoff radius for bonding fracture; $B_1 = -0.55$ and $U_0 = 34.6878$ are constant parameters; varying value of $B_2 = 1.25 - 1.75$ tunes the strength of quartic bonds, with a higher $B_2$ value representing a stronger quartic bond with higher maximum bond force (see **Figure S1** in **Supporting Information (SI)**).[29] The equilibrated polymer melts were then simulated using the combination of WCA and QUARTIC potentials (LJ+QUARTIC) with different bond strengths, and run for a further equilibration period of $2.5 \times 10^5$ in the *NVT* ensemble using a Nosé-Hoover thermostat.[40, 41]

All NEMD simulations of polymer melts under uniaxial extensional flow were carried out in the *NVT* ensemble[40, 41] (in order to mimic the incompressibility of polymer melts), by integrating the atomic version of the SLLOD equations of motion without considering the mass center of polymer chains[42]

$$\begin{cases} \dot{r}_i = \frac{p_i}{m_i} + r_i \cdot \nabla u \\ \dot{p}_i = F_i - p_i \cdot \nabla u - \alpha p_i \end{cases} \quad (4)$$

where $r_i$, $p_i$ and $m_i$ are the position, peculiar momentum and mass of atom $i$ respectively, $F_i$ is the force on atom $i$ and $\alpha$ is the Langevin or Nosé-Hoover thermostat multiplier. The velocity gradient tensor of the SLLOD equations, $\nabla u$, takes the following form under UEF,

$$\nabla u_{UEF} = \begin{bmatrix} -\frac{\dot{\varepsilon}}{2} & 0 & 0 \\ 0 & -\frac{\dot{\varepsilon}}{2} & 0 \\ 0 & 0 & \dot{\varepsilon} \end{bmatrix} \quad (5)$$

The generalized Kraynik-Reinelt (GKR) method developed by Dobson[43, 44] was used to apply periodic boundary conditions under steady-state UEF, as implemented in the UEF package of LAMMPS.[45] Such GKR boundary conditions enable NEMD flows to remap continually without the simulation box becoming too narrow. A wide range of extension rates $\dot{\varepsilon} = 0.00002 - 0.6$ were considered for all polymer melt models with different chain lengths. We note that in this work atomic SLLOD and thermostatting is used, which can introduce unphysical behavior in UEF simulations. In particular, angular streaming velocity can be incorrectly interpreted as thermal energy using atomic thermostats, and applying the SLLOD equations directly to atoms (monomer beads in this work) rather than to molecular centers of mass can induce intramolecular stresses. However, in this case where there is a dense polymer this should be minimal, and applying the alternative, a molecular thermostat and molecular SLLOD, is problematic because the identity and length of the molecules change with time.

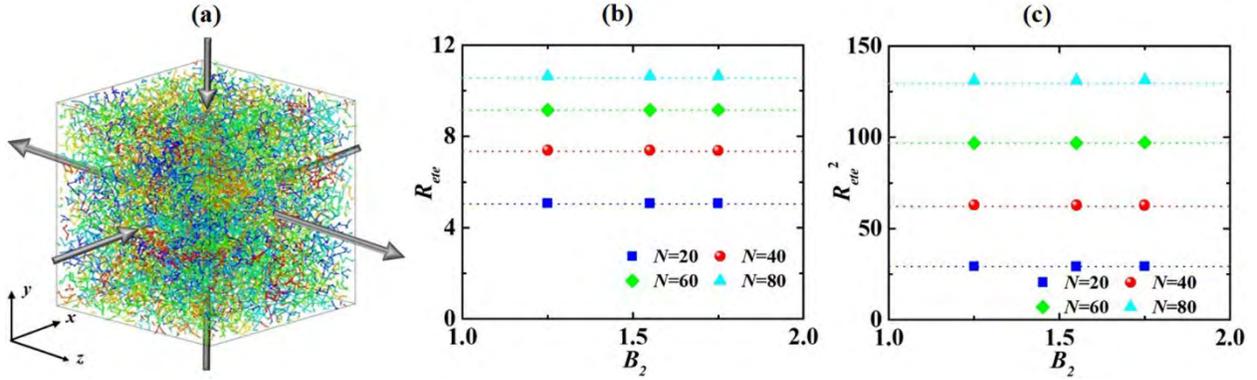

Figure 1. (a) Schematic diagram of the unentangled polymer melt under uniaxial extensional flow. Extension is applied along $z$ axis, and contraction is applied along the $x$ and $y$ axes. (b-c) The predicted average magnitudes of (b) end-to-end vector $\mathbf{R}_{ete}$ and (c) mean-square end-to-end vector $\mathbf{R}_{ete}^2$ of equilibrium polymer melts with chain lengths of $N = 20 - 80$. The scattered symbols are results simulated using the LJ+QUARTIC potential with different strengths of quartic bonds ($B_2 = 1.25 - 1.75$), while dotted lines are results simulated by LJ+FENE potential.

The UEF viscosities of polymer melts with both LJ+FENE and LJ+QUARTIC potentials are calculated at different $\dot{\varepsilon}$, and the chain scission in polymer melts with the LJ+QUARTIC potential is determined at high extension rates. Elongation occurs along the $z$ axis in the UEF simulations, and is accompanied by biaxial contraction along the *x* and *y* axes due to the volume incompressibility, as illustrated in **Figure 1(a)**. The UEF viscosity of the polymer melts can be calculated by the following expression as,

$$\eta_{UEF} = \frac{\sigma_{zz}(\dot{\varepsilon}) - \frac{1}{2}[\sigma_{xx}(\dot{\varepsilon}) + \sigma_{yy}(\dot{\varepsilon})]}{\dot{\varepsilon}} \tag{6}$$

where $\sigma_{ii}(\dot{\varepsilon})$ (*i*=*x*, *y* and *z*) is the flow-induced normal stress, and $\sigma_{zz}(\dot{\varepsilon}) - (\sigma_{xx}(\dot{\varepsilon}) + \sigma_{yy}(\dot{\varepsilon}))/2$ is the normal stress difference. Since the polymer chains do not break, a relatively high time step of $\Delta t = 0.005$ is set in NEMD simulations with LJ+FENE potential at all UEF extension rates. A simulation duration of $3 \times 10^3$ time units is used. In NEMD simulations with LJ+QUARTIC potential, a time step of $\Delta t = 0.005$ is set in NEMD simulations at low and intermediate UEF extension rates for a total simulation duration of $3 \times 10^4$, whereas a time step of $\Delta t = 0.002$ is utilized at high UEF extension rates ($\dot{\varepsilon} > 0.07$) to capture the time-dependent evolution of chain scission under UEF for a simulation duration of $3 \times 10^3$. We tested the effect of time step on the response of $\eta_{UEF}$ with time in the NEMD simulations, and **Figure S2** shows that the time evolution and steady-state values of $\eta_{UEF}$ at high UEF rates are nearly independent of $\Delta t$ in the range of 0.0001 to 0.005 for LJ+FENE potential, as well as in the range of 0.0001 to 0.002 for LJ+QUARTIC potential. Independent NEMD simulations of polymer melts under UEF were run from three different initial configurations, and the averages of these are plotted in the figures unless stated otherwise.

## 3. Results and Discussion

### *3.1. Equilibrium behavior of unentangled polymer melts*

We first study the dynamic structural properties of unentangled polymer melts in their equilibrium states. The average magnitude of the chain end-to-end vector $\langle |\mathbf{R}_{ete}| \rangle$ and its mean-square value $\langle |\mathbf{R}_{ete}^2| \rangle$ are evaluated to provide a quantitative measure of the conformation of

polymer chains, where $\langle \cdots \rangle$ denotes an ensemble average over all polymer chains. In the case of the LJ+FENE potential with either a Langevin or Nosé-Hoover thermostat, the predicted values of $\langle |\mathbf{R}_{ete}| \rangle$ and $\langle |\mathbf{R}_{ete}^2| \rangle$ were very similar to each other, and showed excellent agreement with previous simulation results.[46] For polymer melts with the LJ+QUARTIC potential, it is shown in **Figure 1(b-c)** that the calculated equilibrium structural properties are independent of the strength of the quartic bonds ($B_2$ value), and are close to those results predicted by the LJ+FENE potential. This also reflects the fact that at the temperature considered ($T = 1$), the average kinetic energy is ~ 1.5 per monomer and will be close to the minimum energy and indicating that the bond lengths are close to their equilibrium values in all cases. Furthermore, the bond angle distribution will be determined by non-bonded interactions. Therefore, $\langle |\mathbf{R}_{ete}| \rangle$ and $\langle |\mathbf{R}_{ete}^2| \rangle$ are consistent with each other, and with a classical model where the entropy dominates the distribution of chain length.[47,48] However, the bond strength plays a governing role in rheological behavior and chain scission in polymer melts under UEF, which will be discussed in detail later. The values of $\langle |\mathbf{R}_{ete}^2| \rangle$ for the polymer melt models with both potentials have a linear dependence on (N-1), as expected for an ideal polymer.[49] In addition, the distribution of the magnitude of the chain end-to-end vector $\mathbf{R}_{ete}$ over all chains has a Gaussian-like distribution as shown in **Figure S3**. These structural analyses validate the capability of LJ+QUARTIC potential to reproduce the chain dynamics of unentangled polymer melts.

*3.2. Flow-dependent rheological behavior of unentangled polymer melts*

The rheological properties of unentangled polymer melts were investigated with different chain lengths under a range of UEF extension rates. As shown in **Figure S4-S5**, for both potentials the UEF viscosity $\eta_{UEF}$ gradually increases on commencement of UEF, and approaches a plateau value when the polymer melts reach their steady-state configurations. The averaged plateau value gives the steady-state viscosity ($\eta_{UEF}^{steady}$) under UEF. Irrespective of the selection of potential type and parameters (bond strength $B_2$), the higher the UEF extension rate, the earlier the steady state is reached. The calculated $\eta_{UEF}^{steady}$ of polymer melts with different chain lengths is shown as a function of the UEF extension rate $\dot{\varepsilon}$ in **Figure 2**. It is clear that both LJ+FENE and LJ+QUARTIC potentials exhibit similar trends of $\eta_{UEF}^{steady}$ when $\dot{\varepsilon}$ is lower than certain values in Stage IV. For the lowest values of $\dot{\varepsilon}$ (Stage I in **Figure 2**), $\eta_{UEF}^{steady}$ is approximately constant and independent of extension rate, indicating the Newtonian flow of polymer melts at these values of $\dot{\varepsilon}$.

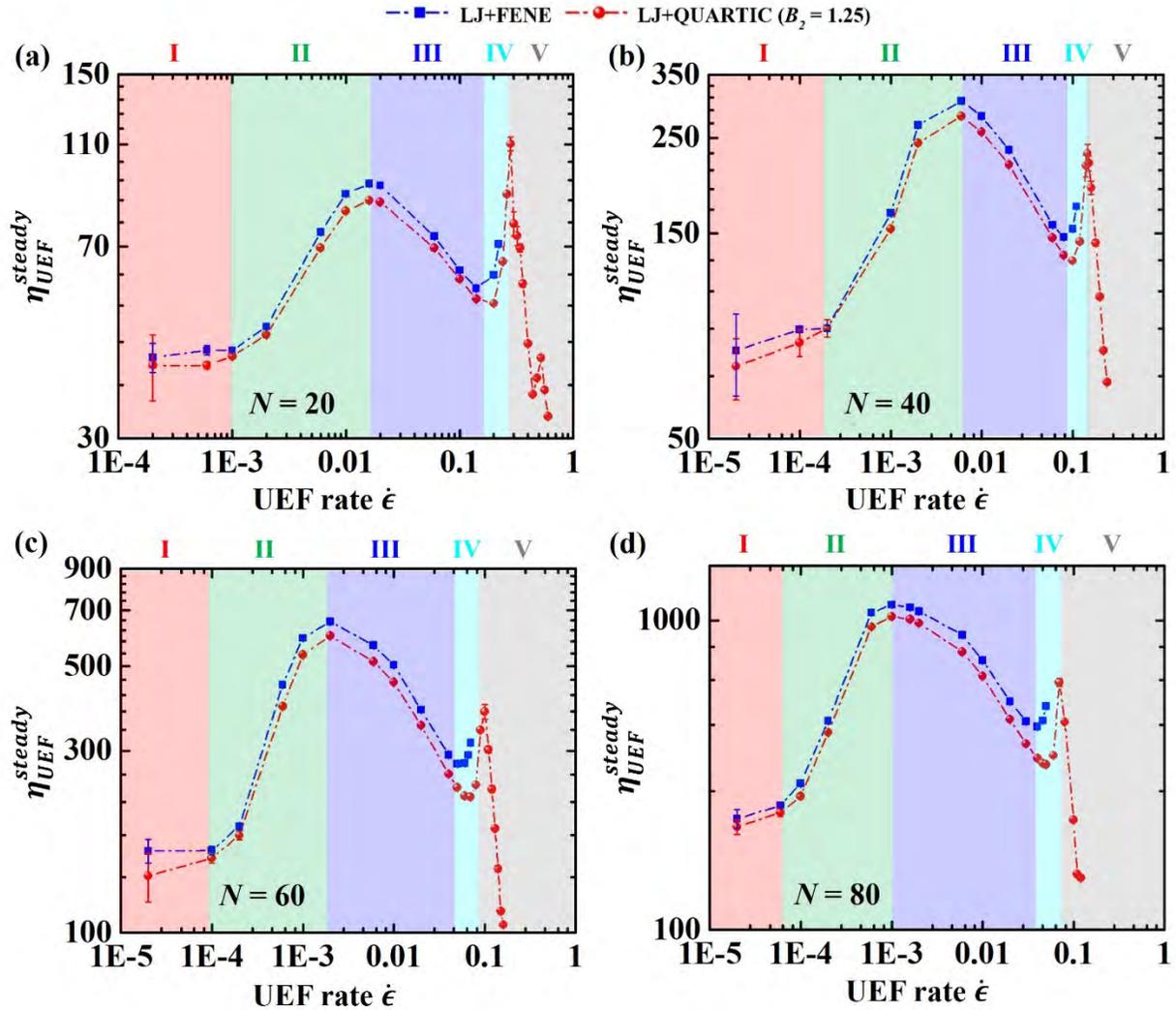

**Figure 2.** The average steady-state values of UEF viscosity $\eta_{UEF}^{steady}$ for unentangled polymer melts with different chain lengths of (a) $N = 20$, (b) $N = 40$, (c) $N = 60$ and (d) $N = 80$, as a function of UEF extension rates, $\dot{\varepsilon}$. Solid circles and squares show the calculated $\eta_{UEF}^{steady}$ using LJ+FENE and LJ+QUARTIC ($B_2 = 1.25$) potentials, respectively. Stages I – V represent the typical UEF-dependent rheological behavior: I – Newtonian; II – Thickening; III – Thinning; IV – Thickening; V – Thinning. The error bars of standard deviation are added in all curves.

In Stage II of the UEF flow, $\eta_{UEF}^{steady}$ increases considerably with an increase in $\dot{\varepsilon}$ (see **Figure 2**), and the behavior using the LJ+FENE and LJ+QUARTIC is similar. Such flow thickening behavior results from the so-called FENE effect due to the initial chain stretch.[50] The error bars of $\eta_{UEF}^{steady}$ also become smaller with the increase of $\dot{\varepsilon}$ due to the faster convergence of viscosity to their steady-state values. To understand the flow-induced evolution of the chain conformation, we evaluate the steady-state values of the chain stretch ratio and an orientation order parameter as a function of $\dot{\varepsilon}$.[51] The chain stretch ratio $\lambda_s$ is defined as $\lambda_s = \langle|\mathbf{R}_{ete}|\rangle / (N-1)l_b$, where $\langle|\mathbf{R}_{ete}|\rangle$ is the average end-to-end distance; $N$ is the number of monomers in the chain; $l_b$ = 0.96 is the equilibrium bond length (minimum energy of the LJ+FENE potential). The orientational order parameter $S_g$ is defined as $S_g = \frac{1}{2}(3\langle\cos^2\theta_g\rangle - 1)$.[23] Here $\theta_g$ quantifies the angle between the director of a polymer compared to the average director of all the polymers, where the director is the eigenvector of the radius of gyration tensor with the largest eigenvalue for the polymer being considered. If all polymers are fully aligned, then $\theta_g = 0$ and $S_g = 1$. **Figure 3** presents results obtained using the LJ+FENE potential and shows that the alignment of the polymers rapidly increases in Stage II ($S_g^{steady} \to 1$), accompanied by partially stretching of the polymer chains ($\lambda_s^{steady} \sim 0.5$) although there is no apparent bond stretch (average bond force $F_{bond}^{steady} < 0$ to balance the non-bonded forces, and constant). The results indicated that the friction between the aligned, partially extended polymers increases in this Stage, increasing the stress and therefore the viscosity. As shown in **Figure S6** in **SI**, the LJ+QUARTIC potential also gives similar evolution of chain conformations. In addition, this relation between chain alignment and shear thickening in Stage II is in contrast to that under pure shear flow, in which case the increase of chain alignment corresponds to the shear thinning behavior of unentangled polymer melts.[23]

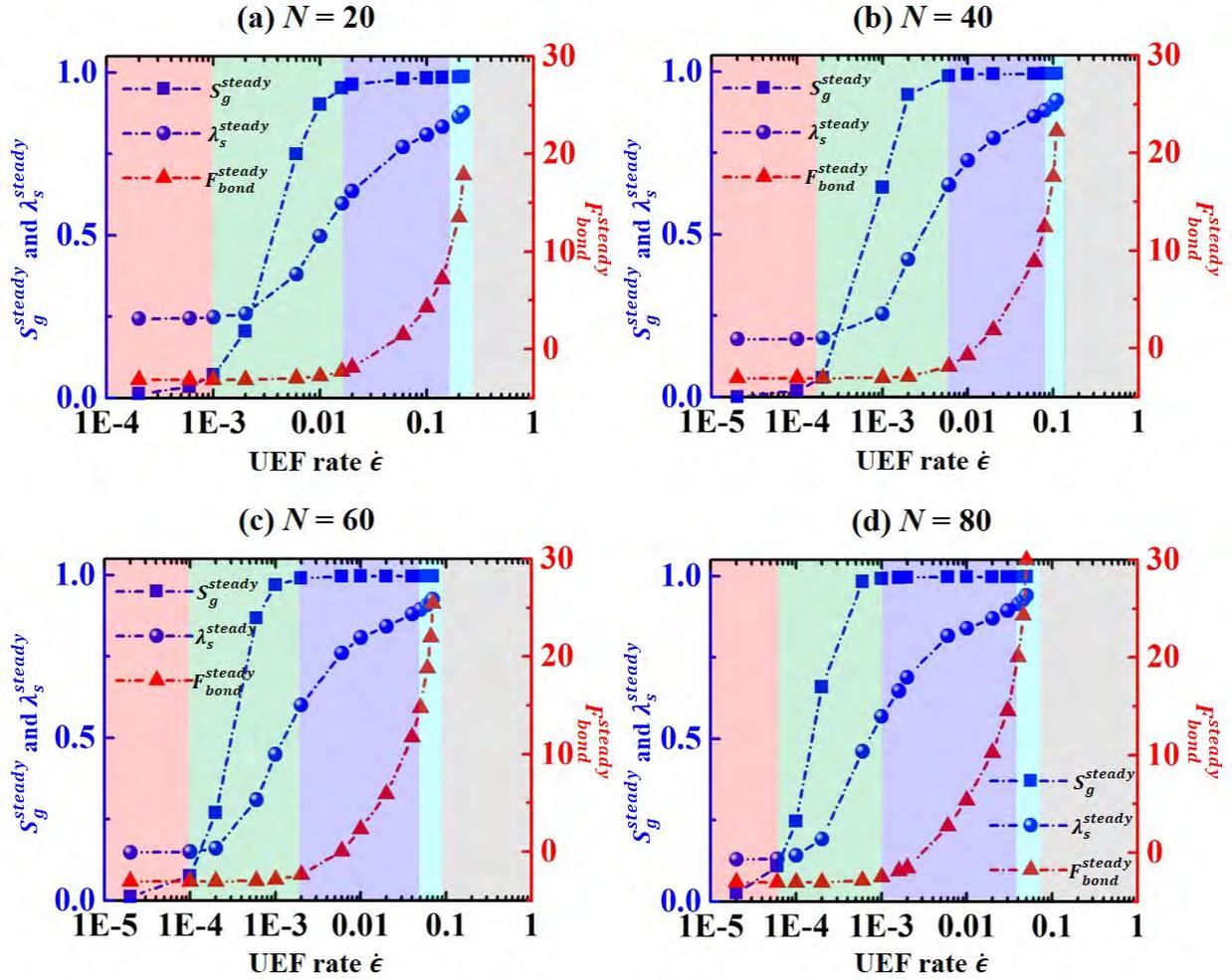

**Figure 3.** The average steady-state values of the order parameter $S_g^{steady}$, chain stretch ratio $\lambda_s^{steady}$ and bond force $F_{bond}^{steady}$ for unentangled polymer melts simulated by the LJ+FENE potential, as a function of UEF extension rates $\dot{\varepsilon}$. Different chain lengths: (a) $N = 20$, (b) $N = 40$, (c) $N = 60$ and (d) $N = 80$ were simulated. The ranges of $\dot{\varepsilon}$ filled with different background colors represent the flow stages illustrated in **Figure 2**.

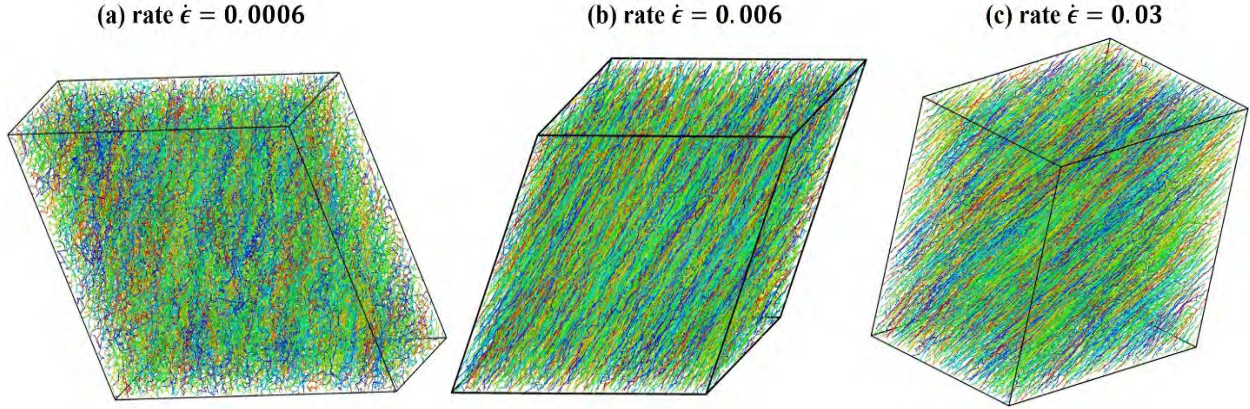

**Figure 4.** The steady-state configurations of polymer melt models at different UEF extension rates of (a) $\dot{\varepsilon} = 0.0006$, (b) $\dot{\varepsilon} = 0.006$ and (c) $\dot{\varepsilon} = 0.03$. The polymer melt models are simulated by LJ+FENE potential, and with the chain length of $N = 80$.

Such flow thickening in Stage II is then followed by a flow thinning associated with the reduction of $\eta_{UEF}^{steady}$ (see Stage III in **Figure 2**). As shown in **Figure 3**, all polymer chains maintain the highly aligned orientation ($S_g^{steady} \sim 1$) during this stage, and the extensional flow also further stretches polymer chains ($\lambda_s^{steady} \to 1$). For example, at $\dot{\varepsilon} = 0.0006$ and $\dot{\varepsilon} = 0.006$ we observe certain degrees of chain alignment and stretching in **Figure 4(a-b)**. This also causes the stretching deformation of some bonds with gradual increase of $F_{bond}^{steady} > 0$. This is consistent with previous studies which show that the alignment suppresses the chain bead friction, due to the high stretching and orientational alignment of the polymer chains under UEF.[51, 52] The thickening-thinning behavior of unentangled polymer melts in Stages II and III also shows an excellent agreement with previous theoretical predictions[53] and experimental data.[50]

With the further increase of $\dot{\varepsilon}$ (Stage IV), the polymers modelled using both potentials have another thickening stage. Our prediction of flow thickening (the increase of $\eta_{UEF}^{steady}$) in Stage IV (see **Figure 2**) shows good agreement with previous simulation work which utilized a molecular formulation of the SLLOD equations of motions.[54] Such interesting thickening behavior can be attributed to the large UEF-induced stretching deformation of bonds, associated with the highly aligned and stretched polymer chains as observed in **Figure 4(c)**. For example, **Figure 3** shows the sharp increase of $F_{bond}^{steady}$ in Stage IV. As a result, the polymer melts can sustain much higher steady-state normal stress difference (see **Eq. 6**), thus leading to the rise of $\eta_{UEF}^{steady}$. In addition, it should be noted that LJ+FENE potentials must fail when $\dot{\varepsilon}$ is above certain critical values, irrespective of the timestep. This is because when the timestep is too long, the corresponding bond forces become infinite, causing the numerical failure of the FENE potential.

A much higher increase of $\eta_{UEF}^{steady}$ is observed during the UEF thickening (Stage IV in **Figure 2**) in the polymer modelled with a LJ+QUARTIC potential than in the polymer modelled with a LJ+FENE potential. This is due to full stretching of QUARTIC bonds without numerical failure like FENE bonds, and consequently the larger increase in $\sigma_{zz}^{steady}$. This results in new thinning behavior at extremely high UEF extension rates (Stage V in **Figure 2**). In this thinning stage of $\eta_{UEF}^{steady}$, UEF triggers the fracture of QUARTIC bonds and shortens the chain lengths of polymer melts. Meanwhile, the error bars of $\eta_{UEF}^{steady}$ become larger again owing to the random bond fracture among polymer chains. As a result, the polydisperse mixture of shorter polymer chains has a much lower $\sigma_{zz}^{steady}$ at the same value of $\dot{\varepsilon}$, giving a reduction in $\eta_{UEF}^{steady}$. This can be supported by our results in **Figure 2** and previous studies[17, 46] that monodisperse (same chain lengths) polymer melts with longer chains are more viscous (higher $\eta_{UEF}^{steady}$) than those with shorter chains.

In addition, we explore the effect of potential on the rheological behavior of polymer melts. It is clearly shown in **Figure 2** that the LJ+FENE potential generally predicts slightly higher values of $\eta_{UEF}^{steady}$ than the LJ+QUARTIC potential does before the initiation of bond fracture. As shown in in **Figure 5**, with the pure LJ+QUARTIC potential, $\eta_{UEF}^{steady}$ is independent of $B_2$ in Stages I-III. This also accords with equilibrium behavior that the configurational entropy governs the elastic behavior of polymer networks at low extension rates.[47, 48] However, in Stage IV and the beginning of Stage V, there is no monotonic trend between $B_2$ and $\eta_{UEF}^{steady}$ due to the complicated behavior of bond fracture and polydispersity of intact polymer chains. In the last range of Stage V, higher $B_2$ corresponds to higher $\eta_{UEF}^{steady}$, which results from fewer broken bonds and will be discussed later.

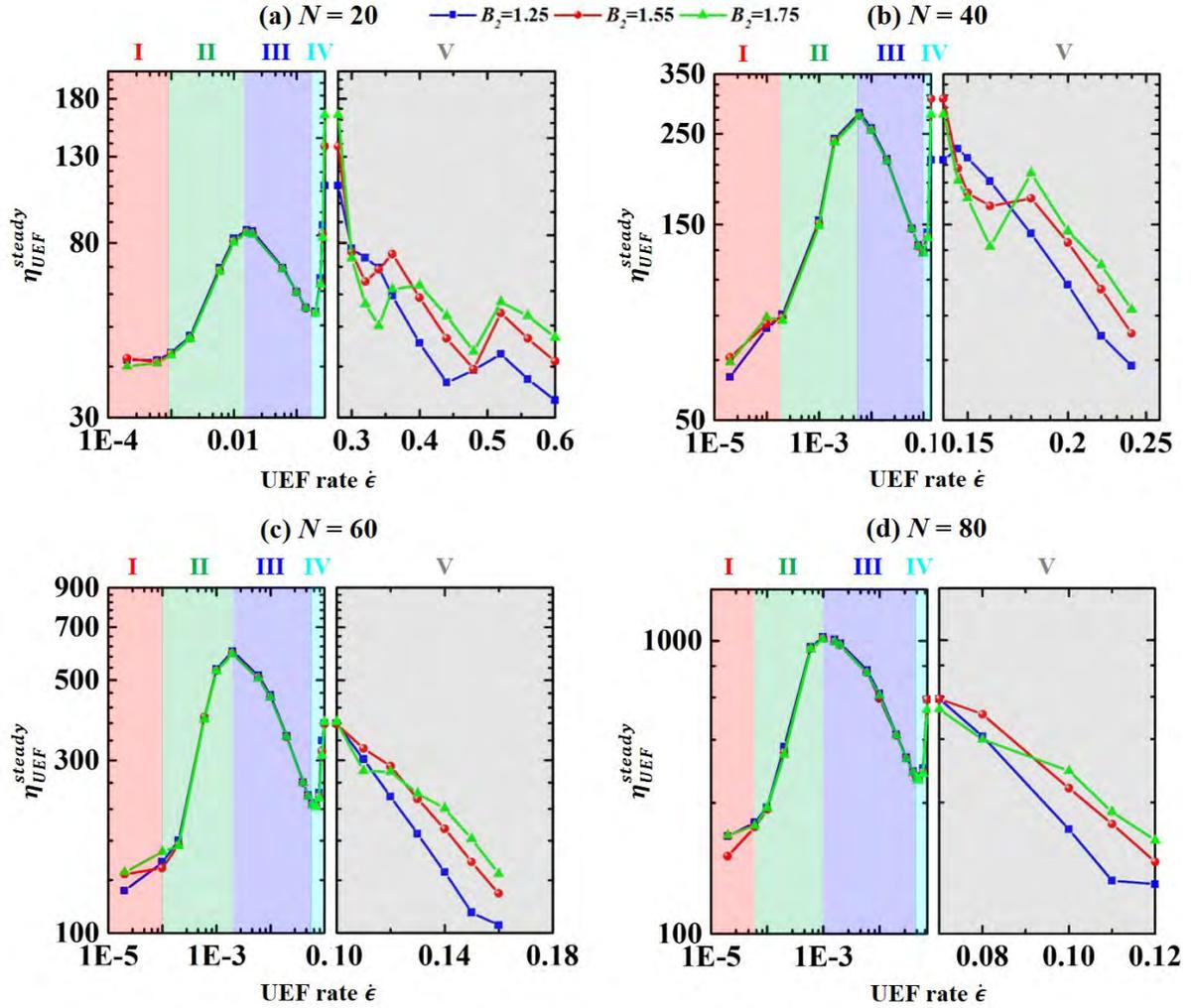

**Figure 5.** The average steady-state values of UEF viscosity $\eta_{UEF}^{steady}$ for unentangled polymer melts with different chain lengths of (a) $N = 20$, (b) $N = 40$, (c) $N = 60$ and (d) $N = 80$, as a function of external UEF rates $\dot{\varepsilon}$. LJ+QUARTIC potentials with different $B_2$ values ($B_2 = 1.25 – 1.75$) are considered, respectively. Stages I – V represent the typical UEF-dependent rheological behavior: I – Newtonian; II – Thickening; III – Thinning; IV – Thickening; V – Thinning.

### 3.3. Flow-induced chain scission in unentangled polymer melts

We finally investigate the chain scission behavior of unentangled polymer melts under UEF. As discussed above, bond fracture occurs when $\dot{\varepsilon}$ reaches a certain critical fracture flow rate $\dot{\varepsilon}_f$. **Figure 6** shows a well-fitted scaling relation between $\dot{\varepsilon}_f$ and the chain molecular weight $N_{chain}$, namely $\dot{\varepsilon}_f \sim N_{chain}^{-1.01}$, which indicates that the scission of longer chains initiates at a lower value of $\dot{\varepsilon}_f$. Such a scaling relation matches that of transient extensional flow $\dot{\varepsilon}_f \sim N_{chain}^{-1}$,[33, 55] rather than for steady-state extensional flow $\dot{\varepsilon}_f \sim N_{chain}^{-2}$.[41, 56] The $\dot{\varepsilon}_f \sim N_{chain}^{-2}$ behavior is expected when the chain is relaxed in a fully extended state before scission occurs, which happens when the scission is slow compared to the full extension of the polymer, and would result in two plateau regions in the time-response curves (see **Figure S5**). The fact that the system does not exhibit this two-step response is consistent with $\dot{\varepsilon}_f \sim N_{chain}^{-1}$. It reflects the fact that in our UEF simulations, a constant extension rate is applied to the equilibrated system at the beginning of the NEMD simulations.

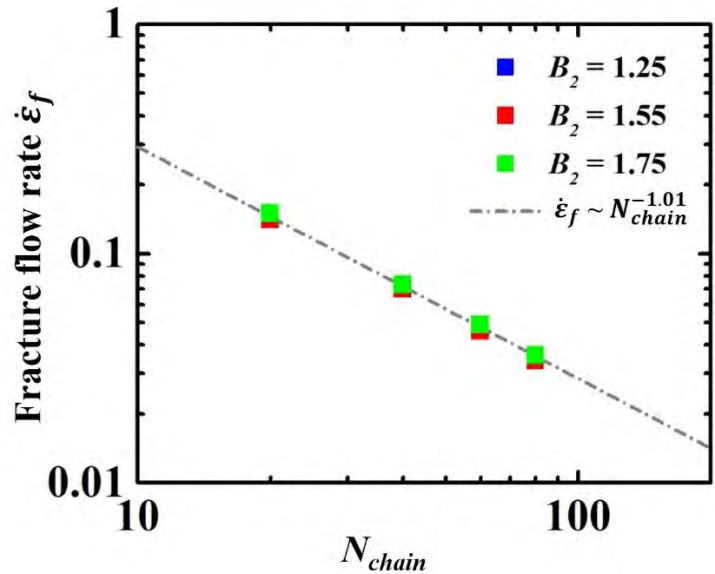

**Figure 6.** Fracture flow rate $\dot{\varepsilon}_f$ as a function of chain molecular weight $N_{chain}$ for LJ+QUARTIC potentials with different $B_2$ values. The dashed line represents the fitted scaling relation of $\dot{\varepsilon}_f \sim N_{chain}^{-1.0}$.

To understand the kinetic behavior of flow-induced bond fracture, we analyze the variation of total number of intact bonds $N_{bond}$ under UEF at different $\dot{\varepsilon}$. As shown in **Figure 7**, the polymer melts possess distinct bond fracture behavior with increasing $\dot{\varepsilon}$. Bond fracture firstly exhibits first-order reaction kinetics, associated with the linear relation between $\ln(N_{bond})$ and total simulation time $t_{NEMD}$ (see the top panel I of all subfigures in **Figure 7**). The slope of the linear region gives the bond-fracture rate constant, $k_f$. As shown in **Figure 8**, and consistent with theoretical work, the rate constant for bond-fracture obeys an Arrhenius relationship. That is, $k_f$ varies exponentially with $\dot{\varepsilon}$, namely $k_f \sim e^{-(E_f - \dot{f}\varepsilon)/k_B T}$ where $E_f$ is the energy barrier for bond break without UEF and $f$ is a constant.[57] This means a higher $\dot{\varepsilon}$ leads to larger $k_f$ as well as more broken bonds. The evolution of chain conformation in **Figure S7** indicates that as $\dot{\varepsilon}$ increases, fewer polymer chains are strongly stretched with lower $\lambda_s$ before bond fracture. A large fraction of bonds stretch, giving higher $F_{bond}$. Thus, bond fracture occurs widely in the polymer network (both at various points within a given polymer and across many polymers) when $F_{bond}$ reaches the maximum forces of QUARTIC bonds.

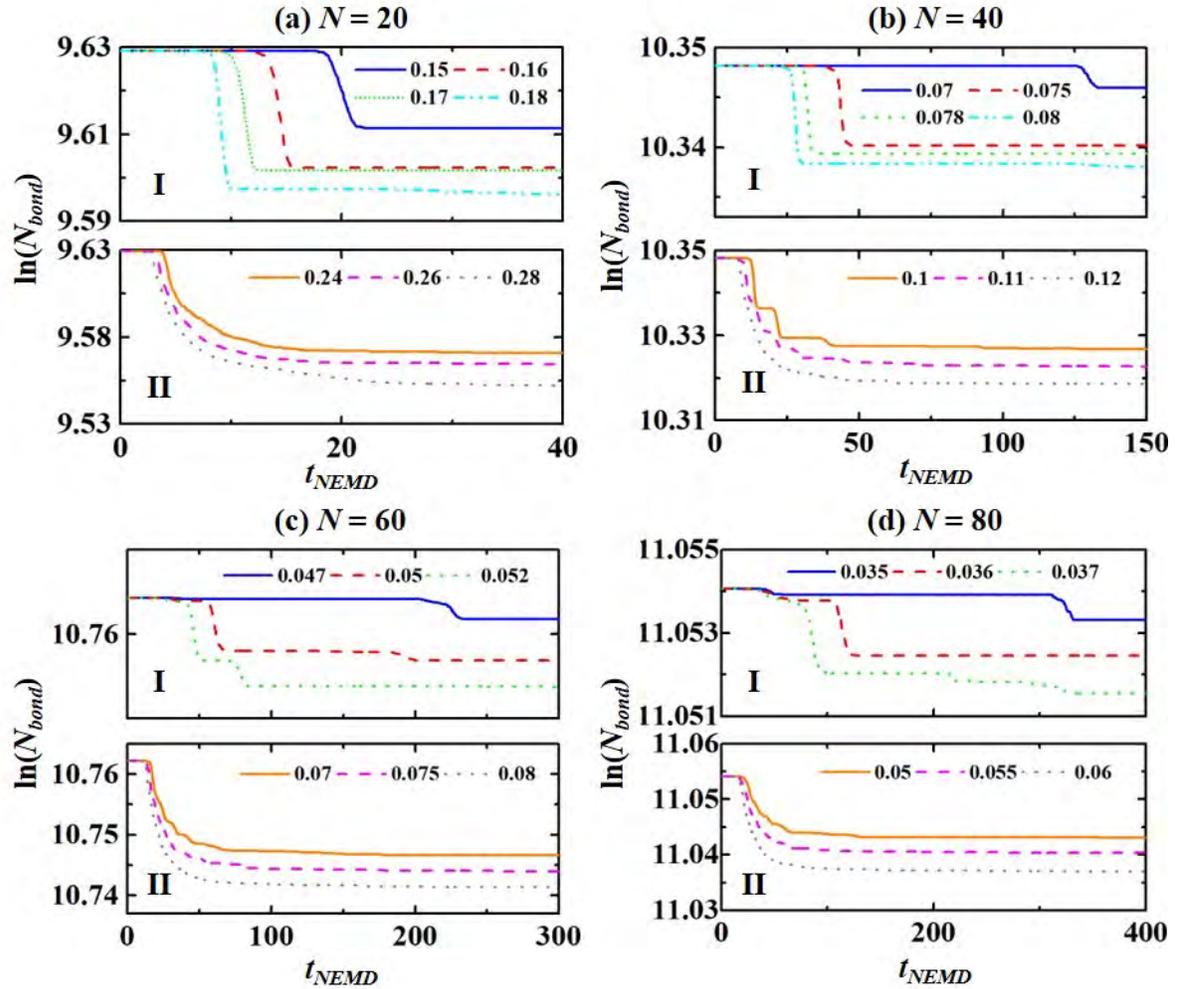

**Figure 7.** Natural logarithm of total number of intact bonds ln($N_{bond}$) for unentangled polymer melts with different chain lengths of (a) $N = 20$, (b) $N = 40$, (c) $N = 60$ and (d) $N = 80$, as a function of total simulation time $t_{NEMD}$ at different UEF extension rates $\dot{\varepsilon}$. The legend gives the values of $\dot{\varepsilon}$. LJ+QUARTIC potential with $B_2 = 1.25$ is considered here. The top panel in each subfigure, I, shows results where the first-order bond fracture kinetics is observed and the bottom panel, II, show shows results at higher extension rate when the kinetics is no longer first-order.

At even higher $\dot{\varepsilon}$, bond fracture no longer exhibits first-order reaction kinetics, as depicted in the bottom panels, II, in **Figure 7**. $N_{bond}$ more gradually decreases with a reduction in the bond-fracture rate. This behavior can be attributed to the fact that extremely high $\dot{\varepsilon}$ results in bond breaking in a polymer occurring before the chain is fully stretched. As a result, polymer melts are only weakly stretched with much lower $\lambda_s$ when bond fracture starts to occur, as shown in **Figure**

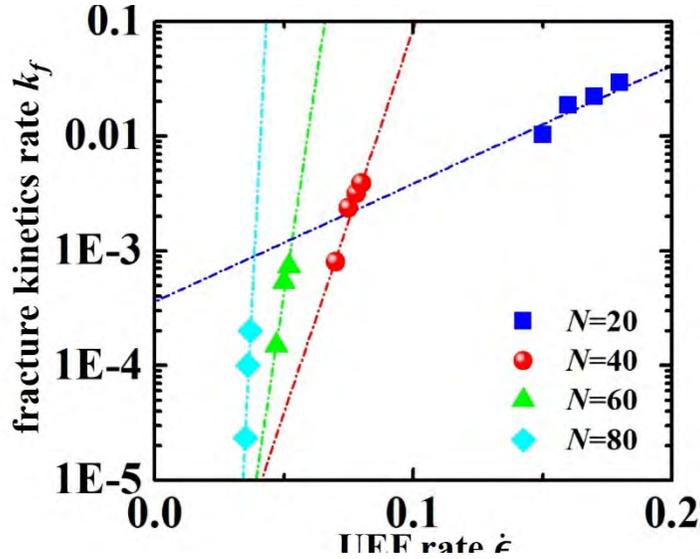

**Figure 8.** Fracture kinetics rate $k_f$ for first-order reaction range as a function of UEF rate $\dot{\varepsilon}$ for unentangled polymer melts with different chain lengths. The dashed lines represent the fitted scaling relation of $k_f \sim e^{-(E_f - \dot{\varepsilon})}$.

**S7**. We note that with higher $B_2$, the force required to fracture a bond is larger, and therefore it will result in slower fracture kinetics with less bond fracture at a given extension rate, and consequently contribute to higher $\eta_{UEF}^{steady}$ (see **Figure 5**). **Figure 7** also shows that there is only a limited reduction in $N_{bond}$. This is because the unentangled polymer melts become polydisperse, with a range of different chain lengths, after the first stage of bond fracture. According to **Figure 6**, the current UEF extension rates are unable to break newly formed shorter chains. Our analysis of flow-induced bond-fracture is also applicable to the activation of mechanophores, which may occur and trigger the functional degradation of manufactured polymers during flow-based manufacturing processes,[1] and will be explored in our future computational work to compare with experimental characterization of flow-induced mechanophore activation in polymer networks.

## 4. Conclusions

In summary, we have performed extensive NEMD simulations to explore the rheological performance and chain scission of unentangled polymer melts under UEF. The effects of chain lengths and UEF extension rates are taken into account, respectively. The predictions of steady-state viscosities demonstrate that unentangled polymer melts exhibit three flow stages of thickening-thinning-thickening. Such flow behavior can be attributed to the evolution of chain conformation, including chain stretching with highly aligned orientation as well as bond stretching under UEF. UEF viscosity then gets highly reduced due to the initiation of bond fracture.

We find that fracture rate $\dot{\varepsilon}_f$ and chain molecular weight $N_{chain}$ follow a scaling relation of $\dot{\varepsilon}_f \sim N_{chain}^{-1.0}$, which is due to the fact that bond fracture is initiated during the transient stage of the extensional flow. In addition, fracture kinetics analysis demonstrates that fracture behavior depends on extension rates. Bond fracture first follows first-order reaction kinetics consistent with high stretching of polymer chains before bond fracture. At higher extension rates, first-order kinetics is no longer observed due to the faster initiation of bond fracture than chain stretching. Our NEMD simulation results not only offer a better understanding of extension-dependent rheological behavior of unentangled polymer networks, but will also provide insightful guidance for the future optimal design of manufacturing processes to avoid undesired mechanical/functional degradation of fabricated polymeric materials. This computational study also indicates interesting avenues of future work on the mixed flow (both shear and extension) behavour of polymeric materials under other processing conditions, such as the free surfaces of electrospinning fibers exposed to atmospheric pressure.

**Acknowledgement**

The authors acknowledge Australian Research Council for its support for this project through the Discovery program (DP190100795). We also acknowledge access to computational resources at the NCI National Facility through the National Computational Merit Allocation Scheme supported by the Australian Government, as well as Pawsey Supercomputing Centre funded by Australian Government and the government of Western Australia.

Supporting Information for

# Mechanical Degradation of Unentangled Polymer Melts under Uniaxial Extensional Flows

Mingchao Wang[1], Stephen Sanderson[1] and Debra J. Searles[*, 1, 2, 3]

[1]Centre for Theoretical and Computational Molecular Science, Australian Institute for Bioengineering and Nanotechnology, The University of Queensland, St Lucia, QLD 4072, Australia

[2]School of Chemistry and Molecular Biosciences, The University of Queensland, St Lucia, QLD 4072, Australia

[3]ARC Centre of Excellence for Green Electrochemical Transformation of Carbon Dioxide, The University of Queensland, St Lucia, QLD 4072, Australia

[*]Corresponding author: email: d.bernhardt@uq.edu.au

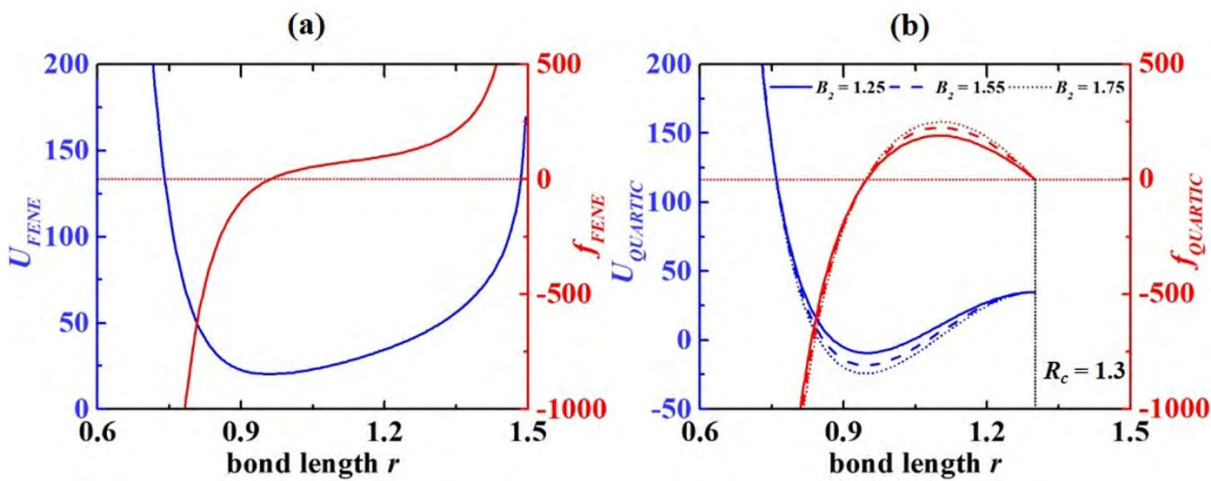

**Figure S1.** Potential energy profiles of atomic interactions for (a) FENE and (b) QUARTIC bonds. Horizontal lines denote the locations of zero bond forces. The solid, dashed and dotted lines in (b) represent the QUARTIC bonds with the bond strength of $B_2$ = 1.25, 1.55 and 1.75, respectively.

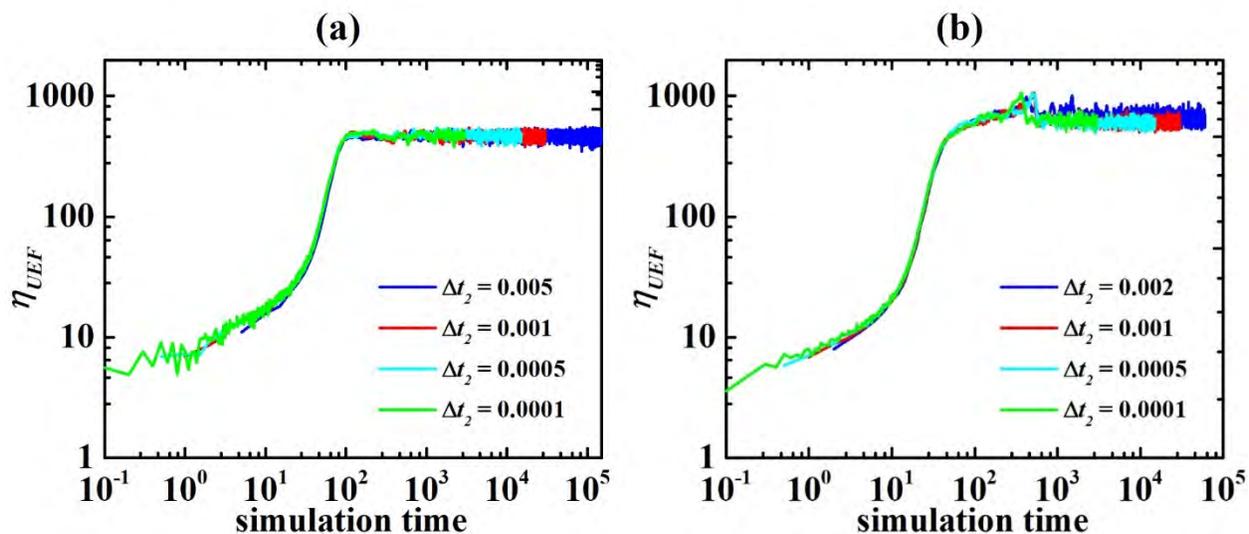

**Figure S2.** The UEF viscosity of unentangled polymer melts ($\eta_{UEF}$) as a function of simulation time at different time steps. The polymers with $N = 80$ beads were simulated using (a) the LJ+FENE potential at a UEF rate of $\dot{\varepsilon} = 0.03$, and (b) the LJ+QUARTIC potential at a UEF rate of $\dot{\varepsilon} = 0.07$, and using Nosé-Hoover thermostat. The initial gradual increase in $\eta_{UEF}$ represents the transient state of the polymer melts, which is followed by a plateau in $\eta_{UEF}$ after $\sim 200$ representing the steady state.

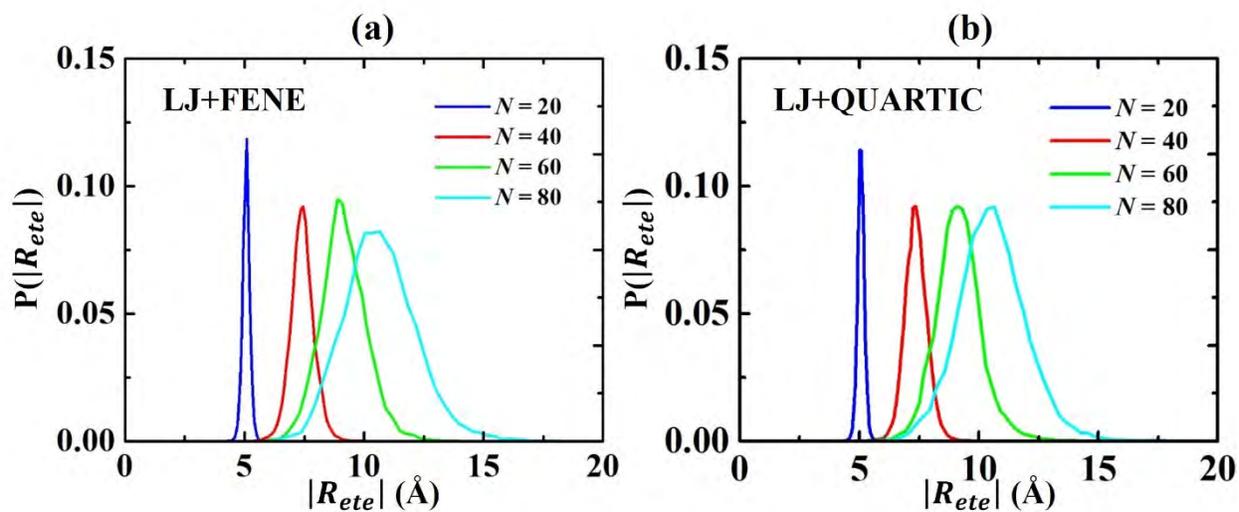

**Figure S3.** The probability distribution of the average magnitude of the chain end-to-end vector $\langle|\mathbf{R}_{ete}|\rangle$ for unentangled polymer melt systems with different chain lengths, as simulated by (a) the LJ+FENE potential and (b) the LJ+QUARTIC potential.

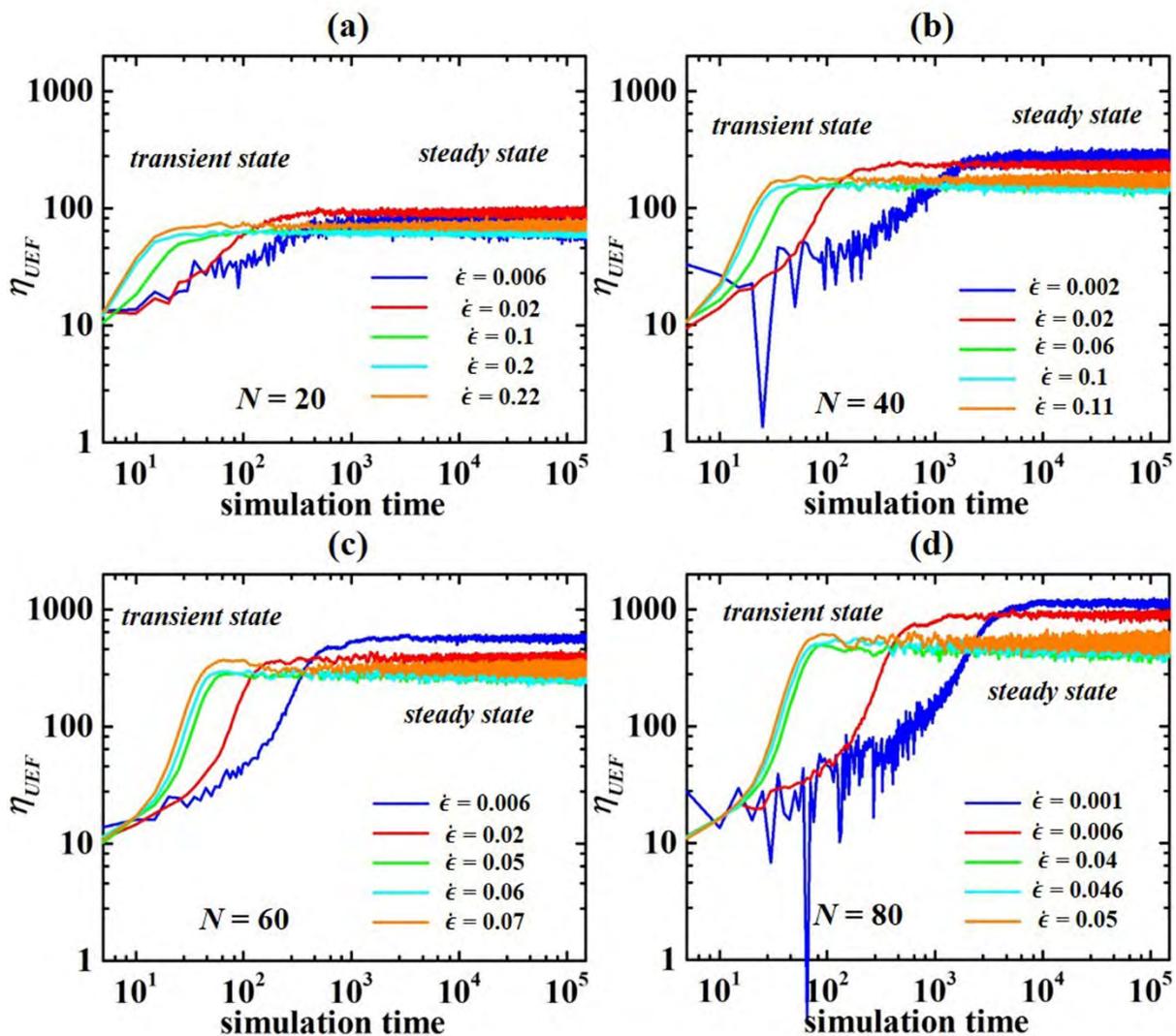

**Figure S4.** The UEF viscosity of the unentangled polymer melts ($\eta_{UEF}$) as a function of time at various UEF extension rates, $\dot{\varepsilon}$. Polymer melts with different chain lengths of (a) $N = 20$, (b) $N = 40$, (c) $N = 60$ and (d) $N = 80$ were simulated using the LJ+FENE potential. The initial gradual increase in $\eta_{UEF}$ is due to the transient states of the polymer melts, and the following plateau in $\eta_{UEF}$ is obtained when a steady state is reached.

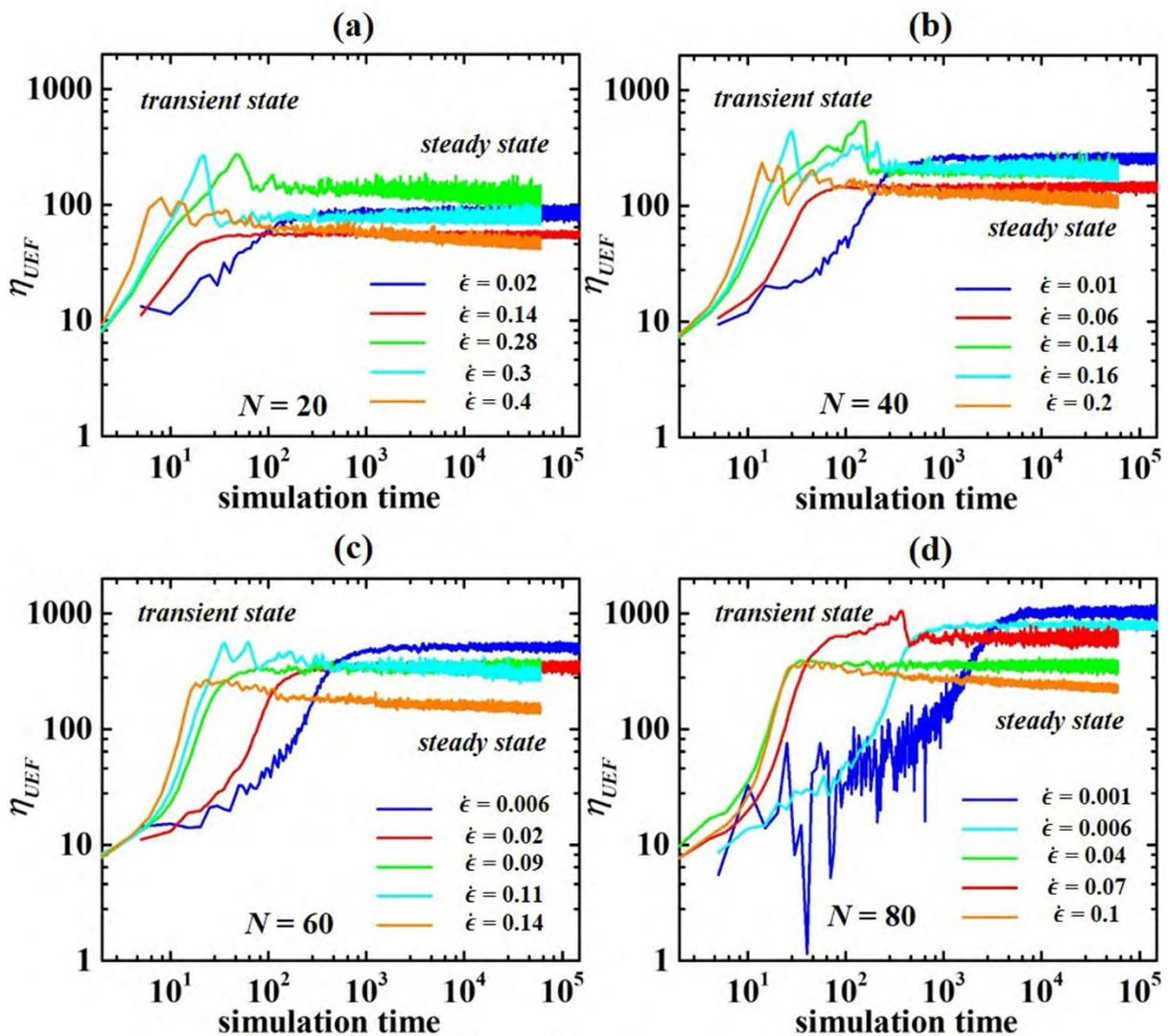

**Figure S5.** The UEF viscosity of the unentangled polymer melts ($\eta_{UEF}$) as a function of time at various UEF extension rates, $\dot{\varepsilon}$. Polymer melts with different chain lengths of (a) $N = 20$, (b) $N = 40$, (c) $N = 60$ and (d) $N = 80$ were simulated using the LJ+QUARTIC potential. The initial gradual increase in $\eta_{UEF}$ is due to the transient states of the polymer melts, and the following plateau in $\eta_{UEF}$ is obtained when a steady state is reached.

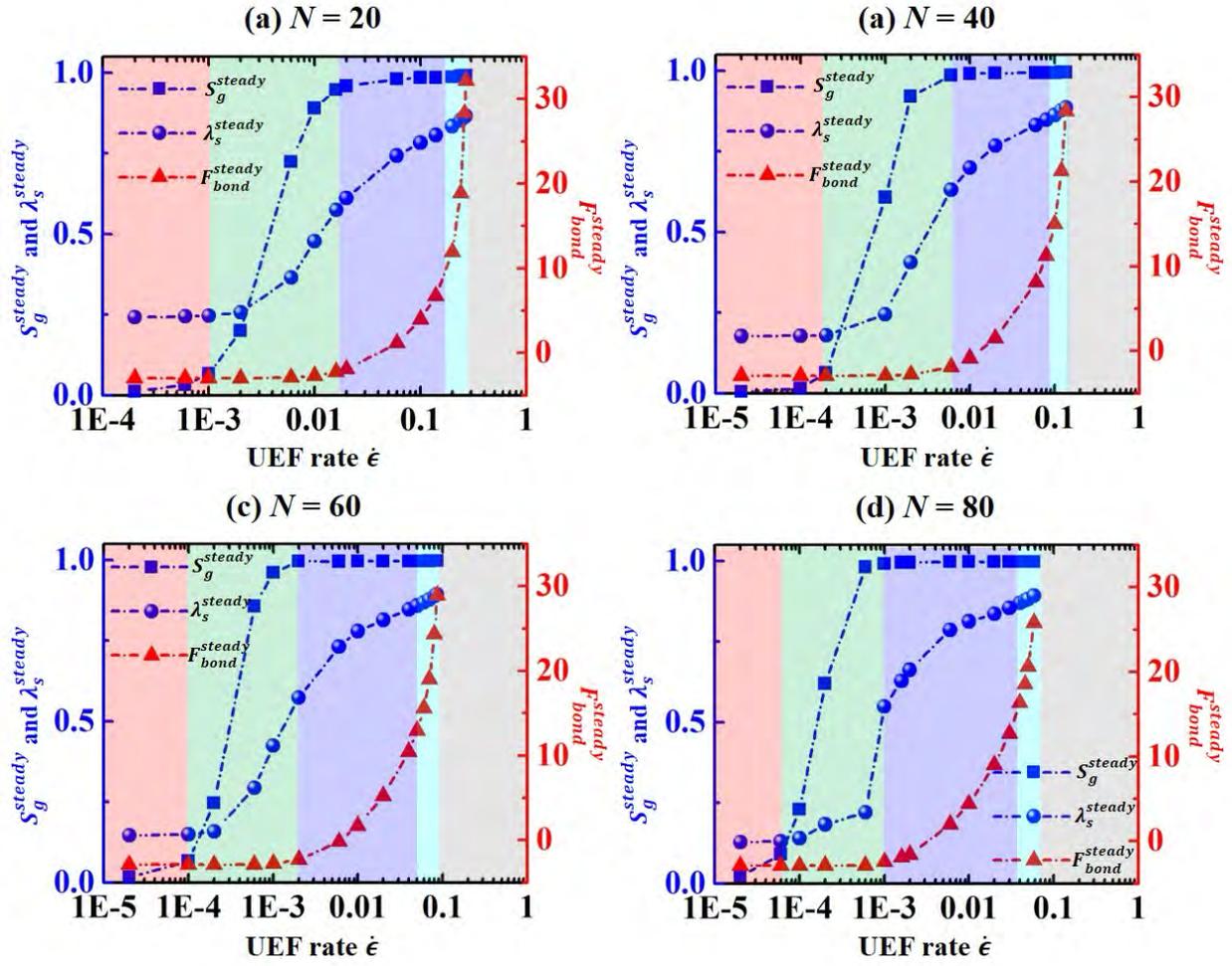

**Figure S6.** The average steady-state values of the order parameter $S_g^{steady}$, chain stretch ratio $\lambda_s^{steady}$ and bond force $F_{bond}^{steady}$ for unentangled polymer melts simulated using the LJ+QUARTIC potential ($B_2 = 1.25$), as a function of UEF extension rates $\dot{\varepsilon}$. Different chain lengths: (a) $N = 20$, (b) $N = 40$, (c) $N = 60$ and (d) $N = 80$ were simulated. The ranges of $\dot{\varepsilon}$ filled with different background colors represent the flow stages illustrated in **Figure 2**.

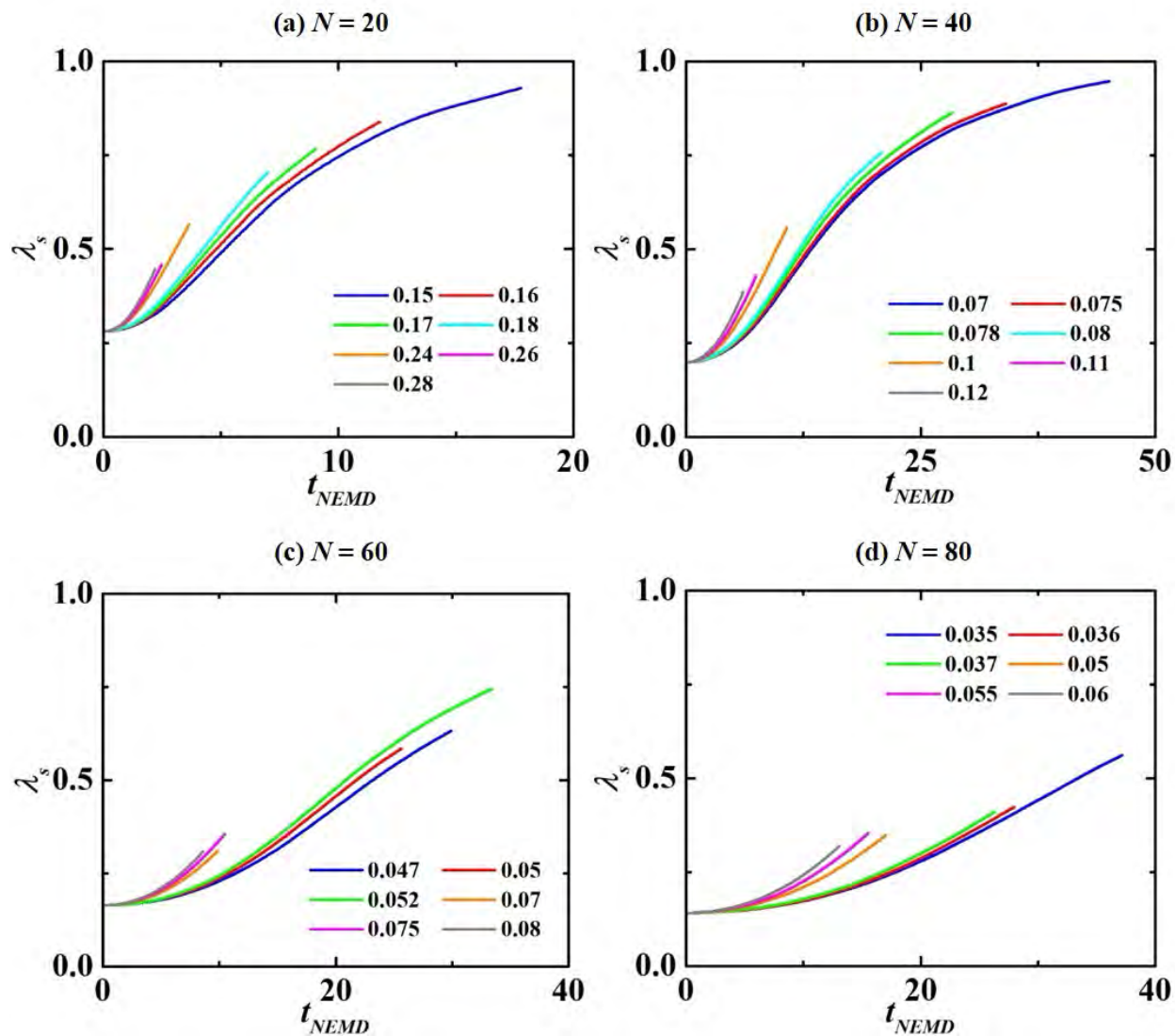

**Figure S7.** Time evolution of the average chain stretch ratio, $\lambda_s$, for unentangled polymer melts with different chain lengths (a) $N = 20$, (b) $N = 40$, (c) $N = 60$ and (d) $N = 80$, after initial application of different UEF extension rates, $\dot{\varepsilon}$. The LJ+QUARTIC potential with $B_2 = 1.25$ is considered here. At the first 3 UEF rates $\dot{\varepsilon}$, bond fracture exhibits first-order reaction kinetics. While at the last 3 UEF rates $\dot{\varepsilon}$, bond fracture does not follow first-order reaction kinetics. The plots are terminated when the first bond is broken.